\documentclass[prc,twocolumn,aps,tightenlines,showpacs,floatfix]{revtex4}
\usepackage{bm}
\usepackage{color}
\usepackage{graphicx,epsfig}
\usepackage{longtable}
\usepackage{multirow}

\usepackage{dcolumn}     

\def\beq{\begin{equation}}
\def\eeq{\end{equation}}
\def\beqy{\begin{eqnarray}}
\def\eeqy{\end{eqnarray}}

\begin{document}

{

\title{Quantum Monte Carlo calculations of electromagnetic
transitions in $^8$Be with meson-exchange currents
derived from chiral effective field theory}

\author{S. Pastore$^1$}
\email{pastores@mailbox.sc.edu}
\author{R. B. Wiringa$^2$}
\email{wiringa@anl.gov}
\author{\mbox{Steven C. Pieper$^2$}}
\email{spieper@anl.gov}
\author{R. Schiavilla$^{3,4}$}
\email{schiavil@jlab.org}
\affiliation{
$^1$Department of Physics and Astronomy, University of South Carolina, Columbia, South Carolina 29208\\
$^2$Physics Division, Argonne National Laboratory, Argonne, Illinois 60439\\
$^3$Theory Center, Jefferson Laboratory, Newport News, Virginia 23606\\
$^4$Department of Physics, Old Dominion University, Norfolk, Virginia 23529
}

\date{\today}

\begin{abstract}

We report quantum Monte Carlo calculations of electromagnetic transitions
in $^8$Be.  
The realistic Argonne $v_{18}$ two-nucleon and Illinois-7 three-nucleon
potentials are used to generate the ground state and nine excited states,
with energies that are in excellent agreement with experiment.
A dozen $M1$ and eight $E2$ transition matrix elements between
these states are then evaluated.
The $E2$ matrix elements are computed only in impulse approximation, with
those transitions from broad resonant states requiring special treatment.
The $M1$ matrix elements include two-body meson-exchange currents derived 
from chiral effective field theory, which typically contribute 20--30\%
of the total expectation value.
Many of the transitions are between isospin-mixed states; the calculations
are performed for isospin-pure states and then combined with
empirical mixing coefficients to compare to experiment.
Alternate mixings are also explored.
In general, we find that transitions between states that have the same
dominant spatial symmetry are in reasonable agreement with experiment, but
those transitions between different spatial symmetries are often
underpredicted.

\end{abstract}


\maketitle

}

\section {Introduction}
\label{sec:intro}

We recently reported {\it ab initio} quantum Monte Carlo (QMC) calculations of 
magnetic moments and electromagnetic (EM) transitions in $A\leq 9$ 
nuclei~\cite{Pastore13}. In that work, the calculated magnetic
moments and $M1$ transitions included corrections arising from
EM two-body meson-exchange currents (MEC) derived in two different approaches:
1) a standard nuclear physics approximation (SNPA)~\cite{MPPSW08,Marcucci05}, 
and 2) the chiral effective theory ($\chi$EFT) formulation of 
Refs.~\cite{Pastore08,PGSVW09,Piarulli12}.
Nuclear wave functions (w.f.'s) were obtained from a Hamiltonian consisting of 
the non-relativistic nucleon kinetic energy plus the Argonne $v_{18}$ (AV18) 
two-nucleon~\cite{WSS95} and Illinois-7 (IL7) three-nucleon~\cite{P08b} 
potentials. 
The SNPA MEC were constructed to obey current conservation with this 
Hamiltonian, while the use of $\chi$EFT MEC constitutes a hybrid calculation.
The two methods are in substantial agreement, producing a theoretical 
microscopic description of nuclear dynamics that successfully reproduces the 
available experimental data, although the $\chi$EFT MEC give somewhat
better results.
Two-body components in the current operators provide significant
corrections to single-nucleon impulse-approximation (IA) calculations.
For example, they contribute up to $\sim 40\%$ of the total predicted value
for the $^9$C magnetic moment~\cite{Pastore13}. 

\begin{figure}
\epsfig{file=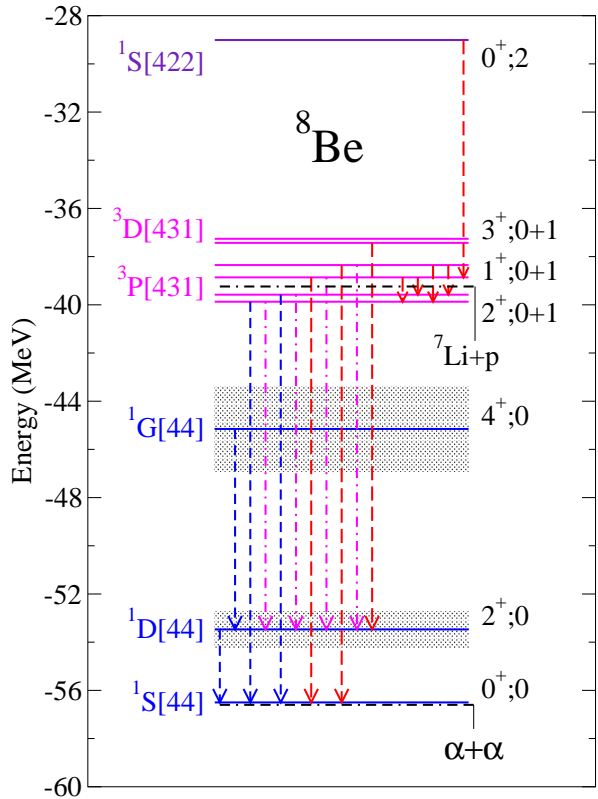,width=9.0cm}
\caption{(Color online) Experimental spectrum of $^8$Be: horizontal lines
denote energy levels, with blue for $T$=0 states, magenta for mixed $T$=0+1 
states, and violet for $T$=2; black dash-dot lines indicate thresholds for 
breakup as indicated and shaded areas denote the large widths of the $^8$Be 
rotational states. 
Vertical lines with arrows indicate the electromagnetic transitions studied:
blue short-dash for $E2$, red long-dash for $M1$, and magenta dash-dot for 
combined $E2$ and $M1$ transitions.}
\label{fig:be8}
\end{figure}

In this work, we implement the framework described above for twenty EM
transitions in the $^8$Be nucleus using only the $\chi$EFT MEC.
The experimental spectrum and EM transitions we consider are illustrated in 
Fig.~\ref{fig:be8}.
This even-even nucleus exhibits a strong two-$\alpha$ cluster structure in 
its ground state,
characterized by angular momentum, parity, and isospin $(J^\pi;T)=(0^+;0)$,
and a Young diagram spatial symmetry that is predominantly [44].
The ground state lies $\sim 0.1$ MeV above the threshold for breakup into
two $\alpha$'s, while the $(2^+;0)$ state at $\sim 3$ MeV excitation and 
$(4^+;0)$ state at $\sim 11$ MeV, are [44] rotational states with large 
($\sim$ 1.5--3.5 MeV) decay widths. 
The next six higher states at 16--19 MeV excitation are three isospin-mixed 
doublets, with the first pair of $(2^+;0+1)$ states lying below the
threshold for breakup into $^7$Li+$p$ and having $\alpha+\alpha$ decay 
widths of $\sim 100$ keV.
The isospin mixing is due to the interplay between $T=0$ states and 
$T=1$ states that are the isobaric analogs of the lowest three states in
$^8$Li and $^8$B, all with the same dominant [431] spatial symmetry.
There are many additional broad excited states above these isospin-mixed 
doublets that are not shown before the final state we consider, the
$(0^+;2)$ isobaric analog of the $^8$He ground state at 27 MeV excitation, 
with dominant [422] spatial symmetry and a very narrow 5 keV decay width.

A comprehensive set of QMC calculations of $A=8$ nuclei was carried out 
in Ref.~\cite{WPCP00} for a Hamiltonian with AV18 and the older Urbana IX 
three-nucleon potential~\cite{PPCPW97}.
More recently, energies, radii, and quadrupole moments of
this nucleus have been recalculated for the [44]
symmetry states~\cite{Datar13}, and for the isospin-mixed states~\cite{WPPM13}, 
using the newer Illinois-7 potential.
The present work complements these studies by calculating many EM
transitions between the low-lying states, which are also illustrated in
Fig.~\ref{fig:be8}.
The $M1$ matrix elements include contributions from two-body $\chi$EFT currents,
which provide important corrections of order 20--30\%.
The two-body current corrections to the $E2$ matrix elements are expected
to be negligible because they appear at higher order in the 
$\chi$EFT expansion~\cite{Piarulli12} and are not computed here.

QMC techniques and $\chi$EFT EM currents were presented in 
Ref.~\cite{Pastore13} and references therein. 
We refer to that work for more details on the calculational scheme
which is here briefly summarized in Sec.~\ref{sec:qmc}.
From there on, we focus on providing and discussing the results. 
In particular, the calculated $^8$Be energy spectrum is presented in 
Sec.~\ref{sec:resenergy}, while results for $E2$ and $M1$ transitions are given 
in Sec.~\ref{sec:resemtrans}. We discuss the results in Sec.~\ref{sec:diss}.

\section {QMC method, Nuclear Hamiltonian and $\chi$EFT EM Currents}
\label{sec:qmc}

EM transition matrix elements are evaluated between w.f.'s
which are solutions of the Schr\"odinger equation:
\begin{equation}
  H \Psi(J^\pi;T,T_z)= E \Psi(J^\pi;T,T_z) \ ,
\end{equation}
where $\Psi(J^\pi;T,T_z)$ is a nuclear w.f.\ with specific spin-parity
$J^\pi$, isospin $T$, and charge state $T_z$. The nuclear Hamiltonian
used in the calculations consists of a kinetic term plus two- and three-body
interaction terms, namely the AV18~\cite{WSS95} and the
IL7~\cite{P08b}, respectively:
\begin{equation}
 H = \sum_i K_i + \sum_{i<j} v_{ij} + \sum_{i<j<k} V_{ijk} \ .
\end{equation}
Nuclear w.f.'s are constructed in two steps. First, a variational Monte
Carlo (VMC) calculation is implemented to construct 
a trial w.f.\ $\Psi_V(J^\pi;T,T_z)$ from products of two- and three-body
correlation operators acting on an antisymmetric single-particle state of
the appropriate quantum numbers.
The correlation operators are designed to reflect the influence of the
interactions at short distances, while appropriate boundary conditions
are imposed at long range~\cite{W91,PPCPW97}.
The $\Psi_V(J^\pi;T,T_z)$ has embedded variational parameters
that are adjusted to minimize the expectation value
\begin{equation}
 E_V = \frac{\langle \Psi_V | H | \Psi_V \rangle}
            {\langle \Psi_V   |   \Psi_V \rangle} \geq E_0 \ ,
\label{eq:expect}
\end{equation}
which is evaluated by Metropolis Monte Carlo integration~\cite{MR2T2}.
Here, $E_0$ is the exact lowest eigenvalue of $H$ for the specified quantum 
numbers.
A good variational trial function has the form
\begin{equation}
   |\Psi_V\rangle =
      {\cal S} \prod_{i<j}^A
      \left[1 + U_{ij} + \sum_{k\neq i,j}^{A}\widetilde{U}^{TNI}_{ijk} \right]
      |\Psi_J\rangle \ ,
\label{eq:psit}
\end{equation}
where the ${\cal S}$ is a symmetrization operator.
The Jastrow w.f. $\Psi_J$ is fully antisymmetric and includes all
possible spatial symmetry states within the $p$-shell that can contribute to the
$(J^\pi;T,T_z)$ quantum numbers of the state of interest, while
$U_{ij}$ and $\widetilde{U}^{TNI}_{ijk}$ are the noncommuting two- and 
three-body correlation operators.

The second step improves on $\Psi_V$ by eliminating excited-state contamination.
This is accomplished by the Green's function Monte Carlo (GFMC) algorithm~\cite{C87-88}
which propagates the Schr\"odinger equation in imaginary time ($\tau$).
The propagated w.f.\ $\Psi(\tau) = e^{-(H-E_0)\tau}\Psi_V$, for large values of
$\tau$, converges to the exact w.f.\ with eigenvalue $E_0$.
In practice, a simplified version $H^\prime$ of the Hamiltonian $H$
is used in the operator, which includes the isoscalar part of the
kinetic energy, a charge-independent eight-operator projection of AV18 called
AV8$^\prime$, a strength-adjusted version of the three-nucleon potential
IL7$^\prime$ (adjusted so that $\langle H^\prime \rangle \sim \langle H \rangle$),
and an isoscalar Coulomb term that integrates to the total charge of the
given nucleus~\cite{KNBSK99}.
The difference between $H$ and $H^\prime$ is calculated using perturbation
theory.
More detail can be found in Refs.~\cite{PPCPW97,WPCP00}.

Matrix elements of the operators of interest are evaluated in terms of a ``mixed''
expectation value between $\Psi_V$ and $\Psi(\tau)$:
\begin{eqnarray}
  \langle O(\tau) \rangle_M & = & \frac{\langle \Psi(\tau) | O |\Psi_V
  \rangle}{\langle \Psi(\tau) | \Psi_V\rangle},
\label{eq:expectation}
\end{eqnarray}
where the operator $O$ acts on the trial function $\Psi_V$.
The desired expectation values, of course, have $\Psi(\tau)$ on both
sides; by writing $\Psi(\tau) = \Psi_V + \delta\Psi(\tau)$  and neglecting
terms of order $[\delta\Psi(\tau)]^2$, we obtain the approximate expression
\begin{eqnarray}
  \langle O (\tau)\rangle &=&
  \frac{\langle\Psi(\tau)| O |\Psi(\tau)\rangle}
  {\langle\Psi(\tau)|\Psi(\tau)\rangle}  \nonumber \\
  &\approx& \langle O (\tau)\rangle_M
    + [\langle O (\tau)\rangle_M - \langle O \rangle_V] ~,
\label{eq:pc_gfmc}
\end{eqnarray}
where $\langle O \rangle_{\rm V}$ is the variational expectation value.

For off-diagonal matrix elements relevant to this work the 
generalized mixed estimate is given by the expression
\begin{eqnarray}
&& \frac{\langle\Psi^f(\tau)| O |\Psi^i(\tau)\rangle}{\sqrt{\langle \Psi^f(\tau) | \Psi^f(\tau)\rangle}
\sqrt{\langle \Psi^i(\tau) |\Psi^i(\tau)\rangle}} \nonumber \\
&\approx&
  \langle O(\tau) \rangle_{M_i}
+ \langle O(\tau) \rangle_{M_f}-\langle O \rangle_V \ ,
\label{eq:extrap}
\end{eqnarray}
where
\begin{eqnarray}
\langle O(\tau) \rangle_{M_f}
& = & \frac{\langle \Psi^f(\tau) | O |\Psi^i_V\rangle}
           {\langle \Psi^f(\tau)|\Psi^f_V\rangle}
      \sqrt{\frac{\langle \Psi^f_V|\Psi^f_V\rangle}
           {\langle \Psi^i_V | \Psi^i_{V}\rangle}} \ ,
\label{eq:mixed_f}
\end{eqnarray}
and $\langle O(\tau) \rangle_{M_i}$ is defined similarly.
For more details see Eqs.~(19--24) and the accompanying discussions in 
Ref.~\cite{PPW07}.
Sources of systematic error in the GFMC evaluation of operator expectation 
values (other than $H^\prime$) include the use of mixed estimates
and the constrained path algorithm for controlling the Fermion sign problem
in the propagation of $\Psi(\tau)$. These are discussed in Ref.~\cite{WPCP00};
the convergence of the current calculations is addressed at the beginning
of Sec.~\ref{sec:resenergy}.

Nuclear EM currents are expressed as an expansion in many-body operators.
The current we use contains up to two-body effects, and is written as:
\begin{equation}
 {\bf j}({\bf q}) =  \sum_i {\bf j}_i({\bf q}) + \sum_{i<j} {\bf j}_{ij}({\bf q}) \ ,
\end{equation}
where ${\bf q}$ is the momentum associated with the external EM field. 
In what follows, we use the notation
\begin{eqnarray}
&&{\bf k}_i={\bf p}_i^\prime-{\bf p}_i \ ,\qquad\qquad {\bf K}_i=\left({\bf p}_i^\prime+{\bf p}_i\right)/2 \ ,\nonumber \\
&&{\bf k}=\left({\bf k}_1-{\bf k}_2\right)/2 \ , \,\qquad {\bf K}= {\bf K}_1+{\bf K}_2 \ ,
\label{eq:ppp}
\end{eqnarray}
where ${\bf p}_i$ (${\bf p}_i^\prime$) is the initial (final) momentum of nucleon $i$, and
${\bf q}={\bf k}_1+{\bf k}_2$ by momentum conservation.

There are two one-body operators resulting from retaining the first two terms
in the $({\bf p}_i/m_N)^2$ expansion of the covariant single-nucleon EM current.
Of course, the leading-order term in this expansion corresponds to the non-relativistic IA
operator consisting of the convection and spin-magnetization single-nucleon currents: 
\begin{equation}
\label{eq:jlo}
{\bf j}^{\rm IA}=\frac{e}{2\, m_N}
\left[ \,2\, e_{N,1} \, {\bf K}_1
+i\,\mu_{N,1}\, {\bm \sigma}_1\times {\bf q }\,\right] \ ,
 \end{equation}
where
\begin{equation}
e_N = (1+\tau_z)/2 \ , \,\,\,
\kappa_N =  (\kappa_S+ \kappa_V \tau_z)/2 \ , \,\,\, \mu_N = e_N+\kappa_N  \ .
\label{eq:ekm}
\end{equation}
Here $\kappa_S= - 0.12$ n.m.\ and $\kappa_V = 3.706$ n.m.\ are the isoscalar (IS) and
isovector (IV) combinations of the anomalous magnetic moments of the proton and neutron,
and $e$ is the electric charge. 

Two-body EM currents are constructed from a $\chi$EFT which retains as explicit degrees of freedom both
pions and nucleons. The resulting operators are expressed as an expansion in nucleon
and pion momenta, generically designated as $Q$.  
The leading-order (LO) contribution in Eq.~(\ref{eq:jlo})
is of order $e\, Q^{-2}$ and contributions up to N3LO or $e\, Q^1$ are retained in the expansion.
These contributions were first calculated by Park {\it et al.} in Ref.~\cite{Park96} using covariant perturbation
theory. More recently, K\"olling and collaborators~\cite{Kolling09-11},
as well as some of the present authors~\cite{Pastore08,PGSVW09,Pastore11,Piarulli12},
derived them using two different implementations of
time-ordered perturbation theory.  In this work, we use the operators 
developed in Refs.~\cite{Pastore08,PGSVW09,Pastore11,Piarulli12}, where
details on the derivation and a complete listing of the formal expressions
may be found. 

The two-body $\chi$EFT EM currents consist of long- and intermediate-range
components described in terms of one-pion exchange (OPE) and two-pion exchange (TPE)
contributions, respectively, as well as contact currents encoding short-range
dynamics. In particular, OPE seagull and pion-in-flight currents appear at 
next-to-leading order (NLO) ($e\, Q^{-1}$) in
the $Q$ expansion, while TPE currents occur at N3LO. The LO and N2LO ($e\, Q^0$)
contributions are given by the single-nucleon operators described above, {\it  i.e.},
the IA operator and its relativistic correction, respectively.

At N3LO, the current operators involve a number of unknown low energy constants
(LECs) which are fixed to experimental data.  The LECs multiplying four-nucleon contact operators
are of two kinds, namely minimal and non-minimal. The former
also enter the $\chi$EFT nucleon-nucleon potential at order $Q^2$ and are therefore fixed by
reproducing the $np$ and $pp$ elastic scattering data, along with the deuteron binding energy.
For these, we take the values resulting from the fitting procedure implemented in Refs.~\cite{Entem03,Machleidt11}.
Non-minimal LECs (there are two of them, one multiplying an isoscalar operator
and the other an isovector operator) need to be fixed to EM observables.

At N3LO, there is also an additional current of one-pion range
which involves three LECs.  One of these multiplies an isoscalar structure,
while the remaining two multiply isovector structures.  As first observed
in Ref.~\cite{Park96}, the isovector component of this current has the same
operator structure as that associated with a $\Delta$-resonance
transition current involving a one-pion exchange. In this type of two-body contribution,
the external photon couples with a nucleon to excite a $\Delta$-resonance state.
The latter decays emitting a pion which is then reabsorbed by a second nucleon.
Given this theoretical insight, one can impose the condition that the two isovector LECs are
in fact given by the couplings of the $\Delta$-resonance current. This mechanism is referred
to as $\Delta$-resonance saturation and has been utilized in various
studies of EM observables of light nuclei (see for
example~\cite{Song07,Song09,Lazauskas11,Piarulli12,Pastore13,Girlanda10}). 
Once the $\Delta$-saturation mechanism is invoked to fix two of the unknown LECs,
the resulting three LECs are fit to the deuteron
and the trinucleon magnetic moments.

The values of the LECs are not unique, in that they depend on the particular momentum cutoff used
to regularize the configuration-space singularities of the EM operators.  In momentum space, these
operators have a power law behavior for large momenta, $k$, which is regularized by
a momentum cutoff of the form $C(k) = {\rm exp}(-k^4/\Lambda^4)$.
For a list of the numerical values of the LECs for $\Lambda=600$ MeV,
which is the cutoff utilized in these calculations, we refer to Ref.~\cite{Pastore13}.  

The N2LO relativistic correction to the one-body IA operator involves two derivatives
acting on the nucleon field. In the GFMC calculation we do not explicitly evaluate
this $ p_i^2$ term, but instead approximate it with its average value, that is
$p_i^2\sim \left< p_i^2\right> $, as determined from the expectation value
of the kinetic energy operator in $^8$Be, from which we obtain
$\left< p_i^2 \right> = 1.375$ fm$^{-2}$. 
This term is a small fraction of the total MEC (see, e.g., 
Table~\ref{tab:vmc-mec} below) so the approximation has little practical effect.

To be consistent with the nomenclature utilized in Ref.~\cite{Pastore13}, we
denote with `MEC' components in the EM currents beyond the IA one-body operator at LO.
However, we stress that the N2LO contribution is a one-body operator,
which does not involve meson-exchange mechanisms.  

\section {$^8$Be Energy Spectrum}
\label{sec:resenergy}

The experimental~\cite{Tilley04} and calculated GFMC energies for the $^8$Be spectrum are 
presented in Table~\ref{tab:energy}, along with the GFMC point proton radii.
The calculations were done by propagating up to some $\tau_{\rm max}$
with an evaluation of observables after every 40 propagation steps, {\it i.e.}, at intervals of
$\tau = 0.02$ MeV$^{-1}$, and averaging in the interval $\tau$=[(0.1 MeV$^{-1})$--$\tau_{\rm max}$];
$\tau_{\rm max}$ is typically  0.3 to 0.4 MeV$^{-1}$.

The calculation of the spectrum is rather involved~\cite{WPCP00},
with two main challenges to face. The first originates from the resonant
nature of the first two excited states (gray shaded states in Fig.~\ref{fig:be8}),
and the ensuing difficulty of extracting a stable resonance energy from
the calculated energies which are evolving to the energy of two separated $\alpha$'s.
This issue was addressed in Ref.~\cite{WPCP00}, and more recently,
however succinctly, in Ref.~\cite{Datar13}. 
The last reference reported an updated measurement of the $E2$ transition 
between the first two excited states of $^8$Be measured via the 
$\alpha+\alpha$ radiative capture with an uncertainty of $\sim 10$\% 
(as opposed to the estimated $\sim 30$\% error of
previous measurements~\cite{Datar05}). 
To accompany the experimental result, a GFMC calculation was performed
for the $E2$ transition matrix element between the
two rotational states, and between the $(2^+;0)$ state and the ground state.
We reprise this calculation in more detail below.

\begin{center}
\begin{table}
\caption{GFMC ground state energy and excitations in MeV for the AV18+IL7
Hamiltonian compared to experiment~\cite{Tilley04} for the $^8$Be spectrum.
Empirical energies are obtained by unfolding the isospin-mixed experimental 
energies using inferred mixing coefficients (see text for explanation). 
Also given are the GFMC point proton (= neutron) radii in fm.
Theoretical or experimental errors $\ge 1$ in the last digit are shown in 
parentheses.}
\begin{tabular}{l >{\hspace{0.8pc}} l >{\hspace{0.8pc}} l  >{\hspace{0.8pc}} l >{\hspace{0.8pc}} l}
\hline
$J^\pi;T$ & GFMC & Empirical & Experiment & $r_p$ \\
\hline
$0^+$     & --56.3(1)  &             &  --56.50       &  2.40 \\
$2^+$     &  + 3.2(2)  &             &  + 3.03(1)    &  2.45(1) \\
$4^+$     &  +11.2(3)  &             &  +11.35(15)   &  2.48(2) \\
$2^+_2;0$ &  +16.8(2)  &  +16.746(3) &  +16.626(3)   &  2.28 \\
$2^+;1$   &  +16.8(2)  &  +16.802(3) &  +16.922(3)   &  2.33 \\
$1^+;1$   &  +17.5(2)  &  +17.66(1)  &  +17.640(1)   &  2.39 \\
$1^+;0$   &  +18.0(2)  &  +18.13(1)  &  +18.150(4)   &  2.36 \\
$3^+;1$   &  +19.4(2)  &  +19.10(3)  &  +19.07(3)    &  2.31 \\
$3^+;0$   &  +19.9(2)  &  +19.21(2)  &  +19.235(10)  &  2.35 \\
$0^+;2$   &  +27.7(2)  &             &  +27.494(2)   &  2.58 \\    
\hline   
\end{tabular}
\label{tab:energy}
\end{table}
\end{center}

The second non-trivial issue is encountered when dealing with the spectrum
of the isospin-mixed states at $16$--$19$ MeV (magenta states in 
Fig.~\ref{fig:be8}). 
These excited states have been extensively discussed in Ref.~\cite{WPPM13}. 
We compute unmixed $T=0$ or $T=1$ states but experimental values are
of course for the mixed states.  The isospin-mixing coefficients can be extracted
from experimental decay widths~\cite{Barker66}.  For the $2^+$ multiplet
this is unambiguous, but for the  $1^+$ and $3^+$ multiplets theoretical
decay widths based on shell-model calculations have been used. 
This is discussed further below. In Table~\ref{tab:energy}
we use the mixing parameters to unfold the ``empirical'' pure-isospin
energies for comparison with our calculations, 
while in subsequent tables we fold the computed EM matrix elements
to generate mixed matrix elements to compare to the data.

We studied the convergence of the GFMC calculations with respect to variations
in the number of unconstrained steps ($n_u$=20 and 50) followed after the
path constraint is relaxed, and found that energies, magnetic moments, and 
rms radii converge at $n_u=20$, which is what is used
for the final results reported here. Most of the calculations
we present are obtained by averaging two calculations, each using 50,000 walkers.
For the physically narrow, nonresonant states, the energy expectation value 
is seen to stabilize at $\tau \sim 0.1$ MeV$^{-1}$.

\begin{figure}
\epsfig{file=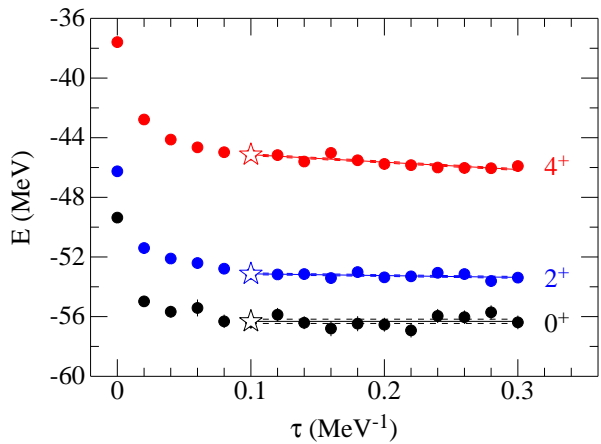,angle=270,width=8.0cm}
\caption{(Color online) GFMC propagation in imaginary time $\tau$ of the 
energy expectation values of the first three states in the $^8$Be spectrum.
Black dots are GFMC propagation points for the ground state, blue dots refer 
to the $(2^+;0)$ rotational state at $\sim 3$ MeV excitation and red dots to the
$(4^+;0)$ state at $\sim 11$ MeV excitation. Solid lines represent a linear 
fit to the GFMC points in the indicated time interval.}
\label{fig:h_res}
\end{figure}

\begin{figure}
\epsfig{file=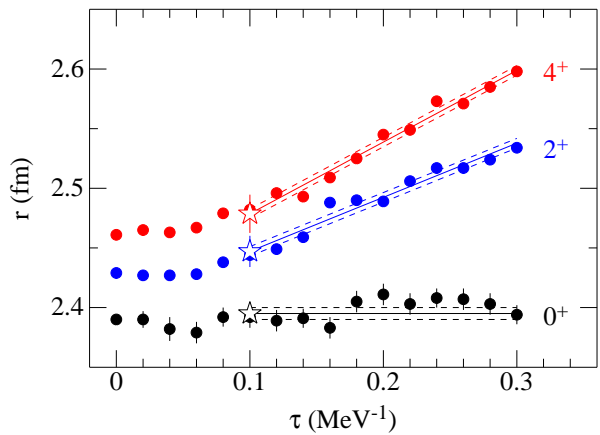,angle=270,width=8.0cm}
\caption{(Color online) GFMC propagation in imaginary time $\tau$
of the point proton radius expectation values of the first three states in
$^8$Be spectrum; notation is the same as in Fig.~\ref{fig:h_res}.}
\label{fig:r_res}
\end{figure}

For the physically wide, resonant states, the binding energy, magnitude of 
the quadrupole moment, and point proton radius all increase monotonically 
as $\tau$ increases.  We interpret this as an indication that the system 
is dissolving into two separated $\alpha$'s. 
In Fig.~\ref{fig:h_res}, we show the GFMC propagation points for the energy 
expectation values of the first three states of $^8$Be. In particular,
the ground state energy is obtained with $n_u=20$ and
20,000 walkers, while the resonant state energies are obtained using $n_u=20$
and averaging two calculations with 50,000 walkers each. From the figure,
we see that the ground state initial VMC energy expectation value at $\tau=0$ 
quickly drops and reaches stability around $\tau=0.1$ MeV$^{-1}$ 
(this point is indicated in the figure with an open star). 
The energies of the two resonant states, instead, keep falling with time:
the $(2^+;0)$ state decreases 0.25 MeV over the interval
$\tau = \left[0.1,0.3\right]$ MeV$^{-1}$,
while the $(4^+;0)$ states falls by 1 MeV.
With this declining energy there is a corresponding increase of the point 
proton radius expectation values, as shown in Fig.~\ref{fig:r_res}
and in the magnitude of the (negative) electric quadrupole moment.

Quantities associated with the resonant states have been calculated
assuming that, also for these states, $\tau\sim 0.1$ MeV$^{-1}$
is the point at which spurious contamination in the nuclear w.f.'s have 
been eliminated by the GFMC propagation. Thus, we make a linear fit to
the 
GFMC values  in the interval $\tau = \left[0.1,0.3\right]$ MeV$^{-1}$,
and extrapolate to $\tau=0.1$ MeV$^{-1}$ for the reported values.
The choice of $\tau=0.1$ MeV$^{-1}$ is somewhat arbitrary. To 
account for this uncertainty we increase the GFMC statistical error by
a systematic error that is obtained by studying the sensitivity of the
results with respect to fitting procedures implemented in two different 
intervals, namely $\tau = \left[0.08,0.3\right]$ MeV$^{-1}$
and $\tau = \left[0.12,0.3\right]$ MeV$^{-1}$, while keeping the
same extrapolating point. 
The total error is represented in the figures by the dashed lines.

For the six states at 16--19 MeV excitation, the GFMC calculations are 
done for pure isospin states of either $T=0$ or $1$. 
The w.f.'s of the isospin-mixed states are written as
\begin{eqnarray}
\label{eq:mixwf}
 \Psi_J^a &=& \alpha_J \Psi_{J,T=0} + \beta_J  \Psi_{J,T=1} , \nonumber \\
 \Psi_J^b &=&  \beta_J \Psi_{J,T=0} - \alpha_J \Psi_{J,T=1} , 
\end{eqnarray}
where the mixing angles satisfy $\alpha_J^2+\beta_J^2=1$.
As one can see from Fig.~\ref{fig:be8} and Table~\ref{tab:energy},
experimentally there are two $J^\pi=2^+$ isospin-mixed states
at 16.626 and 16.922 MeV excitation energies, two  $J^\pi=1^+$
states at 17.64 and 18.15 MeV, and two $J^\pi=3^+$ states at
19.07 and 19.235 MeV. The mixing angles are inferred from 
the experimental values of the decay widths. We follow the
analysis carried out by Barker in Ref.~\cite{Barker66}
and update the experimental widths with more
recent values to obtain the following mixing coefficients~\cite{WPPM13}:
\begin{eqnarray}
\label{eq;alpha_J}
&& \alpha_2 = 0.7705(15), \,\,\,\,\,       \beta_2 = 0.6375(19) \, , \nonumber \\
&& \alpha_1 = 0.21(3),    \,\, \qquad  \beta_1 = 0.98(1) \, , \\
&& \alpha_3 = 0.41(10) ,  \qquad  \beta_3 = 0.91(5) \,  .\nonumber 
\end{eqnarray}
Mixing coefficients for the $2^+$ states are well known because for these states
there is only one decay channel energetically open, that is the 2$\alpha$
emission channel, for which the experimental widths are known with $\sim 0.5$\%
accuracy.
For the other isospin-mixed states, multiple decay channels are available,
which makes the extraction of the mixing coefficients less direct. 
In addition theoretical values of $M1$ matrix elements must be used;
the values above were obtained using traditional shell-model without two-body
current contributions to the matrix elements~\cite{WPPM13,Barker66}.
Revised mixing parameters for the $1^+$ pair, computed using the $M1$ matrix 
elements developed here, are discussed in Sec.~\ref{sec:diss}.

The eigenenergies of the isospin-mixed states, in Table~\ref{tab:energy} 
are given by
\begin{equation}
 E_{a,b}= \frac{H_{00}+H_{11}}{2} \pm \sqrt{\left(\frac{H_{00}-H_{11}}{2}\right)^2 +
\left(H_{01}\right)^2}
\end{equation}
where $H_{00}$ is the diagonal energy expectation in the pure $T$=0 state,
$H_{11}$ is the expectation value in the $T$=1 state, and $H_{01}$ is the 
off-diagonal isospin-mixing (IM) matrix element that connects $T$=0 and 1.
The inferred $H_{00}$ and $H_{11}$ are the empirical values given in 
Table~\ref{tab:energy}.

Finally, the narrow $(0^+;2)$ state at 27 MeV excitation, which has a
dominant [422] spatial symmetry, is a straightforward GFMC calculation.
There could in principle be isospin-mixing with the third $(0^+_3;0)$ state in
the $p$-shell construction of $^8$Be, which also has [422] symmetry, 
via the EM and charge-dependent parts of AV18.  
No such state has been identified experimentally.
A first VMC calculation places this state 0.7(1) MeV higher in excitation
with a 125 keV IM matrix element, which predicts 
$\alpha_0 = 0.19(4)$ and $\beta_0 = 0.98(1)$.
This small amount of mixing may still have a moderate effect on the width of
the physical state, as discussed below.

The overall agreement between experiment and the calculated GFMC spectrum 
for AV18+IL7 shown in Table~\ref{tab:energy} is excellent.  
Only the $3^+$ isospin-mixed doublet is a little too high in excitation and 
a little too spread out compared to the measured values.

\section{Electromagnetic Transitions in $^8$Be}
\label{sec:resemtrans}

We present our results in terms of reduced matrix elements (using Edmonds' 
convention) of the $E2$ and $M1$ operators, the associated $B(E2)$ and $B(M1)$,
and the resulting widths.  For a transition of multipolarity $\lambda$
($X$ designates $E$ or $M$),
\begin{equation}
B(X\lambda)=\left< \Psi_{J_f}\left|\left|X\lambda\right|\right|\Psi_{J_i}\right>^2/(2 J_i+1)
\end{equation} 
is in units of $e^\lambda$ fm$^{2\lambda}$ for electric transitions and 
(n.m.)$^{2\lambda}$ for magnetic transitions.
The widths are given by
\begin{eqnarray}
\label{eq:gammaE}
 \Gamma_{X\lambda} = \frac{8\pi(\lambda+1)}{\lambda[(2\lambda+1)!!]^2} \alpha \hbar c
 \left (\frac{\Delta E}{\hbar c} \right)^{2\lambda+1} B(X\lambda) \ ,
\end{eqnarray}
where $\Delta E$ is the difference in MeV between the experimental initial and
final state energies, $E_i$ and $E_f$; $\alpha$ is the fine-structure constant; and $\hbar c$ is in units of MeV fm.

The calculations of electromagnetic matrix elements 
have been described in detail in Refs.~\cite{PPW07,MPPSW08}. 
Our present results for $E2$ transitions in $^8$Be are given in 
Table~\ref{tab:e2} where the initial and final $(J^\pi;T)$ states and the 
dominant associated spatial symmetries are shown in the first column and
the reduced matrix elements between states of pure isospin are given in the 
second column.
The experimental energies for the physical states are given in the third 
column, and the corresponding theoretical and experimental widths are
shown in the fourth and fifth columns.
We use the IA operator
\begin{equation}
\label{eq:E2-IA}
E2 =  e{\sum_{k} } \frac{1}{2}\left[r_k^2 Y_2(\hat{r}_k)\right](1+\tau_{kz}) 
\end{equation}
without any MEC corrections.

\begin{figure}
\epsfig{file=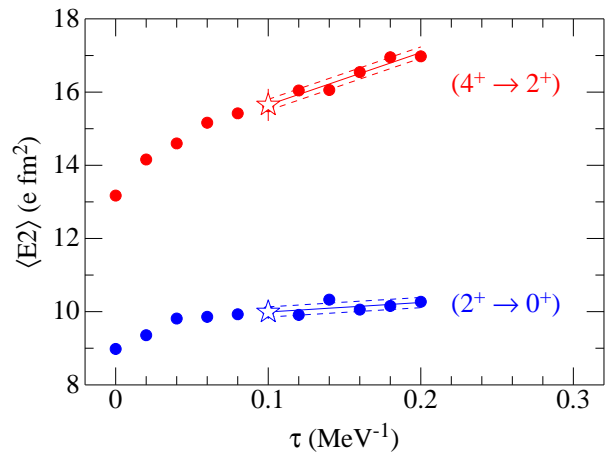,angle=270,width=8.0cm}
\caption{(Color online) GFMC propagation in imaginary time $\tau$ of the 
reduced $E2$ matrix elements among the first three states in $^8$Be spectrum; 
upper red dots are for the $(4^+;0)\rightarrow(2^+;0)$ transition,
lower blue dots are for the $(2^+;0)\rightarrow(0^+;0)$ transition and open
stars denote the extrapolated values.}
\label{fig:e2_res}
\end{figure}

In previous calculations~\cite{PPW07,M+PW12,Pastore13} of nuclei in the
$A=6$--10 range, we have found that $E2$ 
matrix elements of narrow states are generally quite stable under GFMC
propagation, and seldom vary much from the initial VMC estimate.
However, matrix elements from wide states, e.g., for the $^6$Li 
$(2^+;0)\rightarrow(0^+;0)$ decay, show a significant evolution as a
function of $\tau$.
This is also true for the first two transitions in $^8$Be from the broad 
rotational $2^+$ and $4^+$ states.
The matrix element grows monotonically as the GFMC solution evolves in $\tau$
toward a separated $\alpha+\alpha$ configuration, as illustrated in 
Fig.~\ref{fig:e2_res}.
This growth is slow for the lower $(2^+;0)\rightarrow(0^+;0)$ transition, but
much more pronounced for the upper $(4^+;0)\rightarrow(2^+;0)$ transition.
Consequently, we make an extrapolation back to $\tau = 0.1 \pm 0.02$ MeV$^{-1}$
to obtain our best estimate for the matrix element, just as we did for the 
energy and point proton radius discussed above in conjunction with
Figs.~\ref{fig:h_res} and \ref{fig:r_res}.
Our error estimate combines both the Monte Carlo statistical error and the
uncertainty in the extrapolation point.
The numerical results for these two matrix elements and corresponding
decay widths $\Gamma_{E2}$ are reported at the top of Table~\ref{tab:e2}.
The transitions, which are between states of the same dominant [44] spatial 
symmetry, are very large and consistent with a rotor picture for $^8$Be.

\begin{table*}
\begin{center}
\caption{Calculated reduced $E2$ matrix elements and the corresponding decay 
widths compared to experiment~\cite{Tilley04,Datar13}.  The various columns 
show 1) the initial and final $(J^\pi;T)$ states and the dominant associated 
spatial symmetries,
2) the GFMC matrix elements between states of pure isospin, 3) the experimental
energies, and 4) the isospin-mixed theoretical and 5) experimental widths.
In the width values we use the notation $[-x] = 10^{-x}$.}
\label{tab:e2}
\begin{tabular}{c c c c c }
\hline
\hline
$(J^\pi_i;T_i)\rightarrow(J^\pi_f;T_f)$ & \multicolumn{1}{c}{E2[$e$ fm$^2$]}
& $E_i$[MeV]$\rightarrow E_f$[MeV]
& \multicolumn{2}{c}{$\Gamma_{E2}$[eV]} \\
\cline{4-5}
[s.s.]$_i\rightarrow$[s.s.]$_f$  & & &
 \multicolumn{1}{c}{IA} & \multicolumn{1}{c}{Expt.}\\
\hline
\hline
$(2^+;0)\rightarrow(0^+;0)$
   & 10.0(2) &  $3.03 \rightarrow 0.$   & 4.12(16)[-3] & --  \\
$(4^+;0)\rightarrow(2^+;0)$
   & 15.6(4) & $11.35 \rightarrow 3.03$ & 0.87(5)      & 0.67(7) \\
$[44]\rightarrow[44]$  & & & &  \\
\hline
$(2^+_2;0)\rightarrow(0^+;0)$ 
   &  0.55(11)   & $16.626 \rightarrow 0.$ & 1.6(1.0)[--2] & 7.0(2.5)[--2]  \\
$(2^+;1)\rightarrow(0^+;0)$
   &  --0.23(2)  & $16.922 \rightarrow 0.$ & 6.2(2.0)[--2]& 8.4(1.4)[--2] \\
$[431]\rightarrow[44]$  & & & &  \\
\hline
$(2^+_2;0)\rightarrow(2^+;0)$
   &  0.26(7)    & $16.626 \rightarrow 3.03$ & 3.6(2.2)[--3] & --  \\
$(2^+;1)\rightarrow(2^+;0)$
   &  0.03(2)    & $16.922 \rightarrow 3.03$ & 1.7(1.4)[--3]& -- \\
$[431]\rightarrow[44]$  & & & &  \\
\hline
$(1^+;1)\rightarrow(2^+;0)$ 
   &  1.93(6)    & $17.64 \rightarrow 3.03$ &  0.63(5)      & 0.12(5) \\
$(1^+;0)\rightarrow(2^+;0)$
   &  --0.03(5)  & $18.15 \rightarrow 3.03$ &  4.0(1.1)[--2] & -- \\
$[431]\rightarrow[44]$  & & & &  \\
\hline
\hline
\end{tabular}
\end{center}
\end{table*}

We have also calculated an additional six $E2$ transitions
from the isospin-mixed $2^+$ and $1^+$ doublets with dominant [431] spatial 
symmetry, to the $T=0$ ground state or first $2^+$ state.
We denote the isospin-pure matrix elements by
$E2_{T_f T_i} = \left< \Psi_{J_f,T_f}|| E2 ||\Psi_{J_i,T_i}\right>$
and then use the definitions given in Eq.~(\ref{eq:mixwf}) to combine them via
\begin{eqnarray}
 \label{eq:mixme2}
 <\Psi_{J_f,0} || E2 ||\Psi^a_{J_i} > &=&
 \alpha_{J_i} E2_{00} + \beta_{J_i} E2_{01} \ , \nonumber \\
 <\Psi_{J_f,0} || E2 ||\Psi^b_{J_i} > &=&
 \beta_{J_i} E2_{00} -\alpha_{J_i} E2_{01} \ ,
\end{eqnarray}
to evaluate the widths of the physical transitions for comparison to experiment.
Because the $E2$ operator largely preserves spatial symmetry, these transitions
are much weaker than the ones within the $\alpha$-$\alpha$ rotational band.
This makes accurate calculations of these transitions significantly more
difficult.

As an example, we can compare the two $E2$ transitions from the first and
second $2^+$ states to the $0^+$ ground state.
As discussed in Refs.~\cite{PWC04,W06}, the $0^+$ state has five contributing 
$LS$-coupled symmetry components: $^1S[44]$, $^3P[431]$, $^5D[422]$,
$^1S[422]$, and $^3P[4211]$, with the first component having an amplitude
in the present VMC starting w.f. of 0.996.
The $2^+$ states are linear combinations of eight components: $^1D[44]$,
$^3P[431]$, $^3D[431]$, $^3F[431]$, $^5S[422]$, $^5D[422]$, $^1D[422]$, and 
$^3P[4211]$.  The first $2^+$ state also has an amplitude of 0.996 from the
$^1D[44]$ component, while the second $2^+$ state is dominated by the 
$^3P[431]$ component with an amplitude of 0.902.
Consequently, 99\% of the large $E2$ transition from the first excited state 
to the ground state is due to the matrix element between the $^1D[44]$ and 
$^1S[44]$ components.
However, for the much smaller $E2$ transition from the second $2^+$ state, 
this pair of components contributes 1.65 times the final result,
canceled by the matrix element between the two $^3P[431]$ components, which 
gives $-1.44$ times the final result.
The remaining 38 smaller terms, among which there is much additional
cancellation, give 80\% of the total.

Changes in these small components, which may have little effect
on the energy of a given state and hence are not highly constrained by the
GFMC propagation, can have a significant effect on the $E2$ matrix element.
These small components may also be rather sensitive to the three-body
potential in the Hamiltonian, as noted in an earlier study of
$E2$ transitions in $A=10$ nuclei~\cite{M+PW12}. 
This is also true for many of the $M1$ transitions discussed below,
when the initial and final states have different dominant spatial symmetries.

\begin{table*}
\begin{center}
\caption{Calculated reduced $M1$ matrix elements and corresponding decay widths
compared to experiment~\cite{Tilley04}.  The various columns show 1) the
initial and final $(J^\pi;T)$ states and the dominant associated spatial 
symmetries, 2) the GFMC matrix elements between states of pure isospin in IA
and, 3) in total after adding MEC, 4) the \% $z$ of the total given by the 
MEC, 5) the experimental energies, 6) the isospin-mixed theoretical decay 
widths in IA and, 7) in total, and 8) experimental values.
In the width values we use the notation $[-x] = 10^{-x}$.  The results 
marked with a~*~or~\dag~are extra VMC calculations discussed in the text.}
\label{tab:m1}
\begin{tabular}{c l l c c c c c}
\hline
\hline
$(J_i;T_i)\rightarrow(J_f;T_f)$ & \multicolumn{3}{c}{M1[n.m.]}
& $E_i$[MeV]$\rightarrow E_f$[MeV]
& \multicolumn{3}{c}{$\Gamma_{M1}$[eV]} \\
\cline{2-4}\cline{6-8}
[s.s.]$_i\rightarrow$[s.s.]$_f$& \multicolumn{1}{c}{IA}
& \multicolumn{1}{c}{Total}
& \multicolumn{1}{c}{$z$} &
& \multicolumn{1}{c}{IA} & \multicolumn{1}{c}{Total}
& \multicolumn{1}{c}{Expt.}\\
\hline
\hline
$(2^+_2;0)\rightarrow(2^+;0)$  &
0.014(6)  &  0.013(6)  &      & $16.626 \rightarrow 3.03$ & 0.23(3)& 0.51(6)& \\
$(2^+;1)\rightarrow(2^+;0)$  &
0.297(12) &  0.447(18) & 33\% & $16.922 \rightarrow 3.03$ & 0.30(4)& 0.70(7) & \\
  $[431]\rightarrow[44]$ & & & & $16.626 + 16.922 \rightarrow 3.03$  & 0.53(5)  & 1.21(9) & 2.80(18) \\
\hline
$(1^+;1)\rightarrow(0^+;0)$  &
0.551(7)  &  0.767(9)  & 28\% & $17.64 \rightarrow 0.00$ & 6.2(2) & 12.0(3) & 15.0(1.8)\\
$(1^+;1)\rightarrow(2^+;0)$  &
0.398(6)  &  0.567(11) & 30\% & $17.64 \rightarrow 3.03$ & 1.9(1) & 3.8(2) & 6.7(1.3) \\
$(1^+;0)\rightarrow(0^+;0)$  &
0.012(1)  &  0.014(1)  &      & $18.15 \rightarrow 0.00$ & 0.25(1) & 0.50(2) & 1.9(0.4) \\
$(1^+;0)\rightarrow(2^+;0)$  &
0.018(3)  &  0.021(3)  &      & $18.15 \rightarrow 3.03$ & 0.06(1) & 0.13(2) & 4.3(1.2) \\
$[431]\rightarrow[44]$  & & & & & & & \\
\hline
$(1^+;1)\rightarrow(2^+_2;0)$  &
2.287(10) &  2.910(13) & 21\% & $17.64 \rightarrow 16.626$ & 1.92(2)[--2] & 2.97(3)[--2] & 3.2(3)[--2] \\ 
$(1^+;1)\rightarrow(2^+;1)$  &
0.139(2)  &  0.176(3)  & 21\% & $17.64 \rightarrow 16.922$ & 1.22(3)[--3] & 2.20(5)[--3] & 1.3(3)[--3]\\
$(1^+;0)\rightarrow(2^+_2;0)$  &
0.167(3)  &  0.189(3)  & 12\% & $18.15 \rightarrow 16.626$ & 2.52(3)[--2] & 2.87(3)[--2] & 7.7(1.9)[--2]\\ 
$(1^+;0)\rightarrow(2^+;1)$  &
2.596(11) &  2.887(13) & 10\% & $18.15 \rightarrow 16.922$ & 3.26(3)[--2] & 4.18(3)[--2] & 6.2(7)[--2]\\
$[431]\rightarrow[431]$  & & & & &  & & \\
\hline
$(3^+;1)\rightarrow(2^+;0)$  &
0.386(13) &  0.622(22) & 38\% & $19.070 \rightarrow 3.03$ & 0.87(6) & 2.3(2)  & 10.5 \\
$(3^+;0)\rightarrow(2^+;0)$  &
0.015(1)* &  0.030(1)* &      & $19.235\rightarrow 3.03$ & 0.15(2) & 0.37(4) &  --  \\
$[431]\rightarrow[44]$  & & & & & & & \\
\hline
$(0^+;2)\rightarrow(1^+;1)$  &
0.793(7)  &  1.095(8)  & 28\% & $27.49 \rightarrow 17.64$ & 6.7(1) & 12.7(2) & 21.9(3.9) \\
$(0^+_3;0)\rightarrow(1^+;1)$  &
0.553(3)\dag &  0.689(3)\dag & 21\% &               & 8.3(3)\dag& 15.5(5)\dag&           \\
$(0^+_3;0)\rightarrow(1^+;0)$  &
0.073(1)\dag &  0.082(1)\dag & 11\% & $27.49 \rightarrow 18.15$ & 0.28(1)\dag & 0.54(1)\dag & --  \\
$[422]\rightarrow[431]$  & & & & & & & \\
\hline
\hline
\end{tabular}
\end{center}
\end{table*}

An additional complication arises for transitions from the second $2^+$
state because the GFMC propagation is not guaranteed to preserve the
orthogonality of the w.f. relative to the first $2^+$ state.
In practice, GFMC propagation starting from orthogonal VMC w.f.'s
preserves the orthogonality to a high degree~\cite{PWC04};
in this case the amplitude $\langle  \Psi^{2^+_2}(\tau)|\Psi^{2+}_V\rangle$
increases from 0.0010(7) for $\tau$=0 to 0.040(6) averaged over $0.1 \leq \tau \leq 0.3$.
This small admixture leaves the energy and point proton radius of the 
second $2^+$ state as stable functions of
$\tau$, as expected for a narrow state.
However, for the $E2$ matrix element from the $2^+_2$ state to states 
of dominant [44]
symmetry, there are the large cancellations
discussed above and a small admixture of the the $2^+$ state with its large
$E2$ matrix element to states of dominant [44] symmetry can substantially
affect the overlap.  For this reason we have applied a correction by 
orthogonalizing the $\Psi^{2^+_2}(\tau)$ to $\Psi^{2+}_V$,
\begin{equation}
\Psi^{2^{+\prime}_2}(\tau) = \Psi^{2^+_2}(\tau) - 
\langle  \Psi^{2^+_2}(\tau)|\Psi^{2+}_V\rangle \Psi^{2^+}(\tau)  \, .
\end{equation}
This reduces the mixed estimates $\langle \Psi^{2^+_2}(\tau) |E2| \Psi^{2+}_V\rangle$
by 50\% and $\langle \Psi^{2^+_2}(\tau) |E2| \Psi^{0+}_V\rangle$ by 20\%.
This correction is also made for corresponding $M1$ transitions discussed
below, but is relatively much less important.

For the $M1$ transitions
the IA matrix element is evaluated using the $M1$ operator induced by
the one-body current given in Eq.~(\ref{eq:jlo}), namely
\begin{equation}
\label{eq:muIA}
 {\bm \mu}^{\rm IA} =  \sum_{i=1}^A \left(e_{N,i}\, {\bf L}_i + \mu_{N,i}\,
  {\bm \sigma}_i\right)  \ ,
\end{equation}
while the one-body current at N2LO generates the following additional
M1 operator terms~\cite{Pastore08}
\begin{eqnarray}
{\bm \mu}^{\rm N2LO} =&& -\frac{e}{8\, m_N^3}\sum_{i=1}^A 
\Bigg[ \left\{ p_i^2 \, ,\, e_{N,i}\, {\bf L}_i + \mu_{N,i} \, {\bm \sigma}_i\right\} \nonumber \\
&&+ e_{N,i}\, {\bf p}_i\times({\bm \sigma}_i \times {\bf p}_i) \Bigg] \ ,
\label{eq:murc}
\end{eqnarray}
where ${\bf p}_i=-i\nabla_i$ and ${\bf L}_i$ are the linear momentum and
angular momentum operators of particle $i$, and $\{\dots\, ,\, \dots\}$ denotes
the anticommutator.

The matrix element associated with the contribution of two-body currents is
\begin{eqnarray}
\label{eq:muMEC}
&&\langle J^\pi_f ,M_f\mid \! \mu_z^{\rm MEC}\! \mid J^\pi_i,M_i\rangle = \nonumber \\
&& -i \lim_{q\to 0} \frac{2\, m_N}{q}
\langle J^\pi_f,M_f\mid \! j_y^{\rm MEC}(q\, \hat{\bf x})\! \mid J^\pi_i,M_i\rangle \ ,
\end{eqnarray}
where the spin-quantization axis and momentum transfer ${\bf q}$ are, respectively,
along the $\hat{\bf z}$ and $\hat{\bf x}$ axes, and $M_f = M_i$.
The various contributions are evaluated for two small values
of $q<0.02$ fm$^{-1}$ and then extrapolated linearly to the limit $q$=0.
The error due to extrapolation is much smaller than the statistical error
in the Monte Carlo sampling.

In Table~\ref{tab:m1}, we report the results for the $M1$ transition matrix 
elements as well as the decay widths $\Gamma_{M1}$ between the low-lying excited states.
The first column specifies the initial and final states of pure isospin.
The second column, labeled `IA', shows the IA results obtained
with the transition operator of Eq.~(\ref{eq:muIA}), while the third column
labeled with `Total' shows results obtained with the complete EM current 
operator, Eqs.~(\ref{eq:muIA}--\ref{eq:muMEC}).  
The percentage of the total matrix element given by the MEC contributions
is shown in the fourth column.
The fifth column shows the energies of the physical states, while the last
three columns compare the corresponding widths with the 
experimental data from Ref.~\cite{Tilley04}.

\begin{figure*}
\epsfig{file=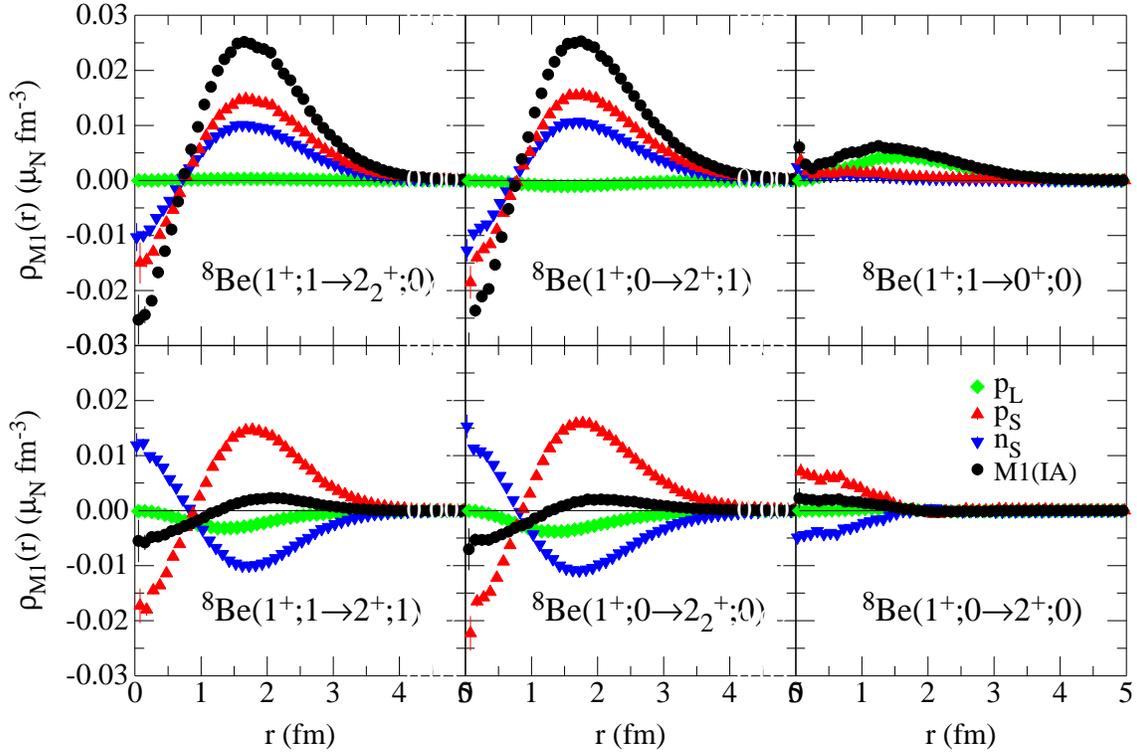,height=6.00in,angle=270}
\caption{(Color online) One-body (IA) $M1$ transition density in nuclear magnetons per fm$^3$ 
for selected M1 transitions (see text for explanation). }
\label{fig:m1_density}
\end{figure*}

As observed in Ref.~\cite{Pastore13}, IA matrix elements are found to have
larger statistical fluctuations than the MEC matrix elements. We separately compute
IA and MEC matrix elements, and then sum the resulting values to obtain the total numbers.

It is worthwhile noting that M1 transitions involving the resonant states
do not monotonically change as $\tau$ increases, a behavior unlike the
quadrupole moments, point proton radii, and energies of these states. 
This stability is understood by observing that the (2$^+$;0) and (4$^+$;0)
rotational states in $^8$Be are $\sim$99\% pure $^1$D$_2$[44] and
$^1$G$_4$[44] states, so they are quantized with L=2 and L=4, respectively.  The orbital
contribution to the magnetic moment is just L/2 nuclear magnetons
because only protons contribute, {\it i.e.}, it is equal to 1.00 n.m.\
in the (2$^+$;0) state and 2.00 n.m.\ in the (4$^+$;0) state.  Because
it is quantized, the magnetic moment should not vary as the
nucleus starts to break up in the GFMC propagation, unlike the
point proton radius where $r$ is growing as $\tau$ increases.
Due to this stability, we can safely propagate M1 matrix elements involving
resonant states to larger values of $\tau$ and average the GFMC result
in larger $\tau$ intervals. 

As for the $E2$ transitions above, the $M1$ matrix elements are evaluated 
between states with well defined isospin, $T=0$ or $1$.
We denote these matrix elements as 
$M1_{T_f T_i} = \left< \Psi_{J_f,T_f}||\mu ||\Psi_{J_i,T_i}\right>$,
with $T_f$ and $T_i$ equal to $0$ or $1$.
For transitions involving isospin-mixing in the initial or final state,
we use expressions similar to Eq.(\ref{eq:mixme2}) to generate the physical
transition rates.
For transitions in which both the initial and final states are isospin-mixed, 
using the definitions given in Eq.~(\ref{eq:mixwf}),
we obtain the following expressions for the isospin-mixed M1 transition matrix elements:
\begin{eqnarray}
\label{eq:mixme1}
 <\Psi^a_{J_f} || M1 ||\Psi^a_{J_i} > &=&
 \alpha_{J_f}\alpha_{J_i} M1_{00} +\alpha_{J_f} \beta_{J_i} M1_{01} \nonumber \\
 &+&\beta_{J_f}\alpha_{J_i}  M1_{10} +\beta_{J_f} \beta_{J_i}  M1_{11} \ , \nonumber\\
 <\Psi^b_{J_f} || M1 ||\Psi^a_{J_i} > &=&
 \beta_{J_f}\alpha_{J_i} M1_{00} +\beta_{J_f} \beta_{J_i} M1_{01} \nonumber \\
 &-&\alpha_{J_f}\alpha_{J_i}  M1_{10} -\alpha_{J_f} \beta_{J_i}  M1_{11} \ , \nonumber\\
 <\Psi^a_{J_f} || M1 ||\Psi^b_{J_i}> &=&
 \alpha_{J_f}\beta_{J_i} M1_{00} -\alpha_{J_f} \alpha_{J_i} M1_{01}   \\
 &+&\beta_{J_f}\beta_{J_i}  M1_{10} -\beta_{J_f} \alpha_{J_i}  M1_{11} \ , \nonumber \\
 <\Psi^b_{J_f} || M1 ||\Psi^b_{J_i} > &=&
 \beta_{J_f}\beta_{J_i} M1_{00} - \beta_{J_f} \alpha_{J_i} M1_{01} \nonumber \\
 &-&\alpha_{J_f}\beta_{J_i}  M1_{10} +\alpha_{J_f} \alpha_{J_i}  M1_{11} \nonumber \ .
\end{eqnarray}
The isospin-mixed M1 matrix elements are used to evaluate the widths as given 
in Eq.~(\ref{eq:gammaE}) for comparison to experiment. The IA and total values
are reported in the sixth and seventh columns of Table~\ref{tab:m1}, and the
experimental widths (where available) are given in the last column of the table.

\begin{figure*}
\epsfig{file=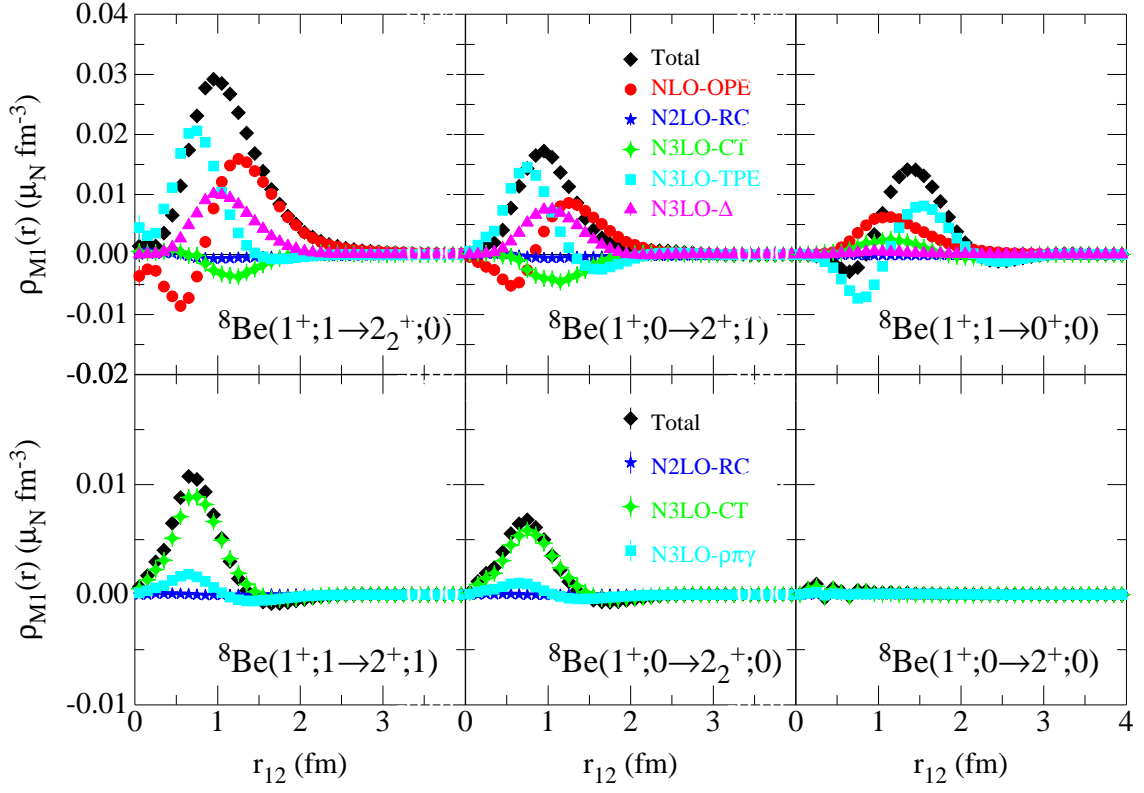,height=6.00in,angle=270}
\caption{(Color online) Two-body (IA) $M1$ transition density in nuclear 
magnetons per fm$^3$ for selected M1 transitions (see text for explanation). }
\label{fig:eft_density}
\end{figure*}

Three extra transitions that were calculated only in VMC are marked by
a~*~or~\dag~in Table~\ref{tab:m1}; they may affect the physical decay widths 
through isospin mixing.
The $(3^+;0)\rightarrow(2^+;0)$ transition marked by a~*~is tiny and
its isospin mixing has little effect on the transition from the physical 
19.07 MeV state.
The corresponding transition from the 19.235 MeV state is predicted to
be much smaller and has not been reported experimentally.
Perhaps more interesting and important, although speculative, is the isospin 
mixing of the proposed $(0^+_3;0)$ state, discussed at the end of 
Sec.~\ref{sec:resenergy}, into the physical 27.49 MeV state, as
shown in the last two lines in Table~\ref{tab:m1} marked by a~\dag.
The line above these gives the result assuming the physical state is pure
$T=2$, and even with MEC contributions, the theoretical width is noticeably
underpredicted.
The first line marked with a~\dag~shows that mixing in the $(0^+_3;0)$ state, 
using $\alpha_0 = 0.19(4)$, increases the decay width 20\%, bringing it
closer to experiment.
The final line shows the corresponding decay to the 18.15 MeV state as
much smaller and thus unlikely to be observed.
A fourth possible transition in this group, $(0^+;2)\rightarrow(1^+;0)$, has
$\Delta T = 2$ and vanishes in IA and also for the MEC considered in this paper.

The $M1$ transitions shown in Table~\ref{tab:m1} can be sorted into four
categories, characterized by having large, medium, small, and tiny matrix 
elements.
The two largest matrix elements are between states of the same spatial symmetry
that change isospin: $(1^+;1)\rightarrow(2^+_2;0)$ and
$(1^+;0)\rightarrow(2^+;1)$.
All four states involved have predominant [431] spatial symmetry, so there is
maximum overlap between the w.f.'s.
Further, because $\Delta T=1$, the spin-magnetization terms of the protons
and neutrons add constructively.
This feature is illustrated in the top left and center panels of
Fig.~\ref{fig:m1_density}, where we plot the IA contributions to the 
magnetic transition density from Eq.(\ref{eq:muIA}), evaluated with the
starting VMC w.f.'s.
In the figure, the red upward-pointing triangles show the proton spin
contribution, the blue downward-pointing triangles show the neutron spin
contribution, the green diamonds are the proton orbital term, and the
black circles give the total transition density.
In both these transitions, the spin contributions are large and the 
proton orbital piece is very small, resulting in a total matrix element of
$\sim 3.0$ n.m.

There are also two transitions between states of the same spatial symmetry
where isospin is conserved, {\it i.e.}, $\Delta T=0$, which results in small matrix
elements: $(1^+;1)\rightarrow(2^+;1)$ and $(1^+;0)\rightarrow(2^+_2;0)$.
These are illustrated in the bottom left and center panels of 
Fig.~\ref{fig:m1_density}.
The magnitudes of the proton spin and neutron spin contributions are very 
similar to the $\Delta T=1$ case, but they have opposite signs and cancel
against each other, and there is a more substantial proton orbital term
which further reduces the total, leading to matrix elements of $\sim 0.2$ n.m.
The values of the VMC densities integrated over $d^3r$ are given in
Table~\ref{tab:vmc-mec} for the transitions shown in the upper and lower 
left panels of Fig.\ref{fig:m1_density}.

Next, there are five matrix elements which are between states of different
spatial symmetry, and are $\Delta T=1$ transitions, such as the 
$(1^+;1)\rightarrow(0^+;0)$ transition illustrated in the top right
panel of Fig.~\ref{fig:m1_density}.
These transitions have proton and neutron spin contributions that add
coherently, but are small because of the small overlap of the initial
and final w.f.'s.
However, they have larger proton orbital pieces, which also add coherently
and dominate the total, leading to medium-size matrix elements in
the range 0.5--1.0 n.m.

\begin{table}
\caption{Individual IA and MEC contributions to one isovector and one isoscalar
M1 transition matrix elements in units of n.m.\ calculated in VMC,
corresponding to the two left-hand panels of 
Figs.~\ref{fig:m1_density} and~\ref{fig:eft_density}.}
\label{tab:vmc-mec}
\begin{tabular}{ldd}
\hline
\hline
$(J_i, T_i)\rightarrow(J_f, T_f)$ 
& \multicolumn{1}{c}{$(1^+;1)\rightarrow(2_2^+;0)$} 
& \multicolumn{1}{c}{$(1^+;1)\rightarrow(2^+;1)$} \\
\hline
IA-$p_L$      &  0.031(1) & -0.224(1) \\
IA-$p_S$      &  1.442(7) &  1.267(7) \\
IA-$n_S$      &  0.988(5) & -0.867(5) \\
IA Total      &  2.461(13)&  0.176(3) \\
\hline
NLO-OPE       &  0.457(1) &           \\
N2LO-RC       & -0.059(1) & -0.001    \\
N3LO-TPE      &  0.090(1) &           \\
N3LO-CT       & -0.038(1) &  0.040(1) \\
N3LO-$\Delta$ &  0.160(1) &           \\
N3LO-$\rho\pi\gamma$ &    & -0.008    \\
MEC Total     &  0.610(2) &  0.031(1) \\
\hline
\hline
\end{tabular}
\end{table}

Finally, there are three matrix elements between states of different
spatial symmetry that have $\Delta T=0$, and these are tiny.
An example is the $(1^+;0)\rightarrow(2^+;0)$ transition in the lower
right panel of Fig.~\ref{fig:m1_density}.
In these cases the proton and neutron spin terms are small in magnitude
and of opposite sign, and the proton orbital piece is also very small,
resulting in matrix elements $< 0.03$ n.m.

The net contribution of MEC EM currents (where MEC = Total - IA) is best 
appreciated by looking at matrix elements between states with well-defined 
isospin, as given in the second to fourth columns of Table~\ref{tab:m1}.
The quantity $z$ in the fourth column is the percentage contribution of
the MEC to the total; it is not given, if the MEC is less than the statistical
error of the total.
MEC contributions to $\Delta T = 0$ transitions are generally smaller than
$\Delta T = 1$ transitions.
This is due to the fact that the major MEC correction, 
given by the OPE seagull and pion-in-flight terms at NLO,
is purely isovector, and cannot contribute to $\Delta T=0$ transitions.
Therefore, only higher order terms, {\it i.e.}, terms at N2LO and N3LO, contribute to
these matrix elements, for which we find $z\sim 10\%$.
Transitions induced by the isovector component of the M1 operator, that is 
transitions in which $T_i\ne T_f$, are instead characterized by a $z$ factor 
spanning the interval $\sim 20-40\%$. 
In general, the NLO currents of one-pion range provide $\sim 60-70\%$ of the 
total MEC correction.
From Table~\ref{tab:m1}, we see that the contribution given by the MEC currents
(with only one exception) improves the IA values, bringing the theory into
better agreement with the experimental data.

It is also interesting to examine the transition magnetic densities due to MEC.
As examples, we discuss the same six transitions whose IA densities are given
in Fig.~\ref{fig:m1_density}.
The associated two-body magnetic densities obtained from MEC terms are shown
in Fig.~\ref{fig:eft_density}, again as calculated with the starting VMC w.f.'s.
For the upper panels, which are isovector transitions,
the red circles labeled `NLO-OPE' show the density due to the long-ranged OPE 
currents, while corrections associated with TPE currents at N3LO are given 
by the cyan squares labeled `N3LO-TPE'. 
Contact current contributions, of both minimal and non-minimal nature,
are represented by the green fort symbols
labeled `N3LO-CT', while the contribution due to the current of one-pion-range, 
which is been saturated by the $\Delta$-resonance, is represented by the 
magenta triangles labeled `N3LO-$\Delta$'. 
In the figure, we also show with blue stars labeled `N2LO-RC' 
the one-body relativistic correction given in Eq.~(\ref{eq:murc}).
The black diamonds give the sum of the various contributions. 
The tail of the magnetic distribution is dominated by the long-range OPE 
contribution, followed by the N3LO-$\Delta$ one; at intermediate- to 
short-range TPE contributions become important. The integrated values
of the individual MEC contributions to the $(1^+;1)\rightarrow(2_2^+;0)$
isovector transition (upper left panel of Fig.~\ref{fig:eft_density})
are listed in Table~\ref{tab:vmc-mec}.

\begin{center}
\begin{table}
\caption{Effect of alternate isospin mixing coefficient $\alpha_1$ 
on $\Gamma_{M1}$; the notation $[-x] = 10^{-x}$.}
\label{tab:imix}
\begin{tabular}{c c c c}
\hline
\hline
$E_i\rightarrow E_f$[MeV] & \multicolumn{3}{c}{$\Gamma_{M1}$[eV]} \\
\cline{2-4}
& \multicolumn{1}{c}{$\alpha_1=0.21$} & \multicolumn{1}{c}{$\alpha_1=0.31$}
& \multicolumn{1}{c}{Expt.}\\
\hline
$17.64 \rightarrow 0.00$   & 12.0(3)      & 11.4(3)      & 15.0(1.8) \\
$17.64 \rightarrow 3.03$   & 3.8(2)       & 3.6(2)       & 6.7(1.3) \\
$18.15 \rightarrow 0.00$   & 0.50(2)      & 1.16(4)      & 1.9(0.4) \\
$18.15 \rightarrow 3.03$   & 0.13(2)      & 0.32(3)      & 4.3(1.2) \\
$[431]\rightarrow[44]$     &              &              & \\
\hline
$17.64 \rightarrow 16.626$ & 2.97(3)[--2] & 3.28(3)[--2] & 3.2(3)[--2] \\
$17.64 \rightarrow 16.922$ & 2.20(5)[--3] & 1.39(4)[--2] & 1.3(3)[--3] \\
$18.15 \rightarrow 16.626$ & 2.87(3)[--2] & 1.84(2)[--2] & 7.7(1.9)[--2] \\
$18.15 \rightarrow 16.922$ & 4.18(3)[--2] & 4.59(3)[--2] & 6.2(7)[--2] \\
$[431]\rightarrow[431]$    &              &              & \\
\hline
\hline
\end{tabular}
\end{table}
\end{center}

Two-body magnetic densities for the isoscalar transitions are shown in
the lower panels of Fig.~\ref{fig:eft_density}.
The isoscalar component of the M1 operator has a rather different structure
in comparison with that of its isovector component;
it has no contributions at NLO, therefore isoscalar transitions are 
suppressed with respect to the isovector ones.
The first correction beyond the IA picture enters at N2LO and is given by
the one-body relativistic correction of Eq.~(\ref{eq:murc}), shown by the 
blue stars labeled with `N2LO-RC'. 
There are two isoscalar contributions at N3LO. 
The first is associated with the tree-level current of one-pion range 
represented by the cyan squares labeled `N3LO-$\rho\pi\gamma$'. 
This isoscalar tree-level current can, in principle, be saturated by the 
$\rho\pi\gamma$ transition current~\cite{Pastore11}, however, we fix its 
associated LEC so as to reproduce the magnetic moments of the
deuteron and the isoscalar combination of the trinucleon magnetic moments,
as explained in Sec.~\ref{sec:qmc}.  
The second is the contact current at N3LO, shown by the green fort 
symbols labeled `N3LO-CT', and these, in fact, dominate the total isoscalar
two-body MEC contribution shown by the black diamonds.
The integrated values for the $(1^+;1)\rightarrow(2^+;1)$ transition
(lower left panel of Fig.~\ref{fig:eft_density}) are also given in 
Table~\ref{tab:vmc-mec}.

\section {Discussion}
\label{sec:diss}

The spatial symmetry-conserving $M1$ transitions are between the isospin-mixed $2^+$ 
and $1^+$ doublets, so the comparison with experimental widths requires 
both the matrix elements between isospin-pure states and the $\alpha_J$ and
$\beta_J$ parameters of Eqs.~(\ref{eq:mixwf},\ref{eq;alpha_J}) as input.
We consider the $\alpha_2$ and $\beta_2$ to be well-determined by the 
$\Gamma_\alpha$ measurements for the $2^+$ doublet.
However, the $\alpha_1$ and $\beta_1$ were first estimated by 
Barker~\cite{Barker66} by looking at the ratio of the $\Gamma_{M1}$'s for 
the $1^+$ doublet and comparing to shell-model calculations.
Instead, we could use our more sophisticated calculations to determine the best isospin-mixing
parameters.

If we minimize the $\chi^2$ with respect to experiment for the four spatial 
symmetry-conserving transitions, i.e., those given in the third block of Table~\ref{tab:m1},
we find $\alpha_1$ = 0.31(4), compared to the ``experimental" value of 0.21(3)
used above in Table~\ref{tab:m1} and discussed in Ref.~\cite{WPPM13}.
The predicted widths for these two isospin-mixing parameters are compared in 
Table~\ref{tab:imix}, along with the four other symmetry-changing
transitions from the $1^+$ doublet to the ground or first excited state;
the $\chi^2$ comparison with experiment for these cases is also improved.
However, this alternate value for $\alpha_1$ implies a significantly larger
IM matrix element $H_{01} = -150(18)$ keV, compared to the
theoretical value for this Hamiltonian of $-94(1)$ keV calculated in
Ref.~\cite{WPPM13}, which was in good agreement with the earlier empirical 
value of $-103(14)$ keV.

The results of our QMC calculations are in fair agreement with experiment
when the transitions are between states of the same spatial symmetry.
However, when the spatial symmetry of the initial and final states is
different, we generally underpredict the reported experimental widths.
The $E2$ calculations of Table~\ref{tab:e2} give large matrix elements
for the $[44]\rightarrow[44]$ transitions and show reasonable agreement with
the recently remeasured $(4^+;0)\rightarrow(2^+;0)$ width.
The calculations underpredict the $[44]\rightarrow[431]$
transitions from the isospin-mixed $2^+$ doublet to the ground state, 
although here both theory and experiment have large error bars.
The predicted transitions to the first $2^+$ are smaller, and perhaps not 
surprisingly unobserved to date.
For the $E2$ transition from the first $1^+$ at 17.64 MeV, we significantly
overpredict the width, due to a surprisingly large $\Delta T = 1$ matrix
element between $^1D[44]$ and $^1P[431]$ symmetry components.
The unobserved transition from the $1^+$ state at 18.15 MeV 
is tiny, due to a vanishing $\Delta T = 0$ matrix element.  
The larger value of $\alpha_1$ discussed above would reduce the discrepancy
with experiment slightly.

The QMC results for $M1$ matrix elements are similar, in that the four 
symmetry-conserving $[431]\rightarrow[431]$ transitions are in fair
agreement with experiment, once MEC contributions are included.
The agreement can be improved further by searching for better isospin-mixing 
parameters, $\alpha_J$ and $\beta_J$, as discussed above.
Seven of the eight symmetry-changing $M1$ transitions are 
underpredicted by amounts ranging from only 25\% to factors of 2--5.
The worst matrix element is the same $(1^+;0)\rightarrow(2^+;0)$ 
transition that also vanishes in $E2$, leading to a decay width for the
18.15 MeV state which is 15--30 times too small.  

Even though many of the experimental widths considered in this work have 
large errors, the serious
discrepancies between some of the experimental and calculated values
highlight the challenge for theory to accurately predict transition amplitudes
between states with dominant admixtures of different spatial symmetry 
or between states consisting of linear combinations of components of different
spatial symmetry and occurring with similar probabilities. 

To our knowledge, Refs.~\cite{Pastore13,MPPSW08} and the present work
are the only {\it ab initio} calculations of EM transitions in 
$A > 4$ nuclei that include MEC contributions.
We find that the calculated $M1$ matrix elements 
have significant contributions, typically at the 20-30\% level, from two-body 
EM current operators, especially from those of one-pion range.
The sizable MEC corrections are found to almost always improve
the IA results for $M1$ transitions. 
This corroborates the importance of many-body effects in nuclear systems, 
and indicates that an understanding of low-energy EM transitions requires 
contributions from MEC in combination with a complete treatment of nuclear 
dynamics based on Hamiltonians that include two- and three-nucleon forces.

\acknowledgments

The many-body calculations were performed on the parallel computers of the 
Laboratory Computing Resource Center, Argonne National Laboratory.
This work is supported by the National Science Foundation,
Grant No. PHY-1068305 (S.P.), and by the U.S.~Department of Energy, Office of Nuclear 
Physics, under contracts No.~DE-FG02-09ER41621 (S.P.), No.~DE-AC02-06CH11357 (S.C.P. and R.B.W.) 
and No.~DE-AC05-06OR23177 (R.S.), and under the NUCLEI SciDAC-3 grant.

\end{document}